\documentclass[preprint,prd,nofootinbib,preprintnumbers]{revtex4}
\usepackage{graphicx, bm}

\begin{document}
%\preprint{UMD-PP-01/15}
\title{ Supersymmetry and R-parity: an Overview\footnote{Invited article for the Richard Arnowitt memorial volume, to be published in Physica Scripta}}

\author{Rabindra N. Mohapatra}
\affiliation{Maryland Center for Fundamental Physics and Department of Physics, University of Maryland, College Park, Maryland 20742, USA}

\begin{abstract} This article provides a brief overview of some of the theoretical aspects of R-parity violation (RPV) in the minimal supersymmetric standard model (MSSM) and its extensions. Both spontaneous and explicit RPV models are discussed and some consequences are outlined. In particular, it is emphasized  that the simplest supersymmetric theories based on local B-L predict that R-parity must be a broken symmetry, a fact which makes a compelling case for taking R-parity breaking seriously in discussions of supersymmetry phenomenology. 
\end{abstract}

\maketitle

\section{Introduction}
In 1982, Richard Arnowitt co-authored a fundamental paper on supergravity with Ali Chamsheddine and Pran Nath, which discussed how to apply supergravity to particle physics
~\cite{ACN,BFS,ACN1} using the formulation of applied supergravity in ~\cite{cremmer,ACN2}. A great deal of theoretical works in subsequent years that applied supersymmetry to particle physics, used the techniques outlined in these papers. Many proposed experimental tests of supersymmetry were proposed in both colliders and low energy processes~\cite{haber1}. The fact that supersymmetry provided a resolution of the hierarchy problem of the standard model added momentum to this research and made it one of the primary areas of activity in beyond the standard model physics. The particular model under the ``microscope" was the minimal supersymmetric extension of the standard model (MSSM).  When the Large Hadron Collider at CERN  was being built, search for supersymmetric partners of the SM particles became one of its primary goals. 

In the most popular version of MSSM, a new symmetry called R-parity (see below) is assumed to be an exact symmetry. This has the implication that the lightest superpartner is absolutely stable and can be identified as the dark matter of the universe. This additional attractive property of MSSM prompted most of the LHC searches for supersymmetry to focus on R-parity conserving modes, which always involve missing energy final states. So far these experimental efforts have not proved successful in uncovering any evidence for supersymmetry, pushing the mass limits on super-partners to uncomfortably large values. The search for supersymmetry will of course continue as soon as the 14 TeV beam turns on.  The absence of evidence for it in the run I, however , has inspired efforts to go beyond the MSSM to include new effects such as R-parity breaking (RPV), extra SM singlet fields, extra gauge symmetries etc. The goal of this paper is to provide a brief overview of  extensions of MSSM that involve R-parity breaking in various theoretical settings and their implications. Its goal is not to provide an exhaustive account of the vast field of RPV models but rather a brief overview of some of salient features of such models. There already exist many excellent reviews of the subject and they should be consulted for detailed implications as well as references.

Application of supersymmetry to particle physics starts with a supersymmetric version of the standard model (SM), where all the fields, both fermionic and bosonic, get ``promoted" to become chiral superfields which have both a bosonic and fermionic parts. The partners of known SM fermionic fields are called ``sfermions" (e.g. squarks, sleptons etc.). Partners of Higgs fields are called ``Higgsinos". The gauge fields become the vector superfields, again with  bosonic components identified with the familiar $W^\pm, Z, \gamma$ whereas their fermionic parts being new fields called gauginos. %There will be much more thorough discussions of MSSM in other articles of this volume as well as other places (see e.g. PDG, 2014~\cite{haber}). 
%I only give what is the usually considered MSSM superpotential :
Most discussions of MSSM are based on the superpotential
\begin{eqnarray}
W_{MSSM}~=~h_uQH_uu^c+h_dQH_dd^c+h_eLH_de^c+\mu H_uH_d
\end{eqnarray}
where $Q, u^c,d^c,L, e^c$ are the matter superfields and $H_{u,d}$ are the up and down type Higgs superfields. To this one must add  the gauge interaction part to get the full supersymmetric Lagrangian. Since supersymmetry is not an exact symmetry of nature, one must also add the supersymmetry breaking terms, which consist of mass terms for superpartners, trilinear terms of type present in the above superpotential but will all terms being the scalar components of the superfields and mass terms for Gauginos, all of which must respect the gauge symmetry.

The terms in the superpotential of Eq. (1)  are however not the only $SU(2)_L\times U(1)_Y$  invariant and supersymmetric terms that could have been written down given the superfields of MSSM. This is where the story of R-parity comes in. It was pointed out in 1978~\cite{FF}, that in models with the above superpotential~(Eq (1)), one can define an exact symmetry of the Lagrangian given by $(-1)^R=(-1)^{3B-L+2S}\equiv (-1)^{3(B-L)+2S}$, where $B$ is baryon number, $L$ is lepton number and $S$ is the spin of the particle. This symmetry is known as R-parity. Under this symmetry, all superpartners of SM fields are odd while all the SM fields are even. R-parity invariance is, of course, of great experimental relevance for collider searches for supersymmetry since it implies that all superpartner fields must be produced in pairs and must decay to states, at least one of which must be an RP-odd particle (or a superpartner field). Also it implies that the lightest superpartner field (LSP) must be absolutely stable. This fact has two cosmological implications: (i)  it implies that if the  LSP is electrically charged, then the theory is cosmologically disfavored~\cite{CHAMPS}; (ii) on the other hand if the LSP is electrically neutral, it can play the role of dark matter of the universe, as eluded to above. These facts are all well known and likely will be discussed in other articles in this volume. The goal of this article is to discuss theories which depart from the above simple paradigm and involve terms that break R-parity before or after gauge symmetry breaking.

R-parity breaking is a vast topic (for a review, see~\cite{rev}) and in this article, no attempt will be made to give comprehensive discussion of all issues and implications of the models. Rather, I give a brief overview of various ways that R-parity breaking emerges, and give examples of some models where R-parity remains a natural symmetry (i.e. the symmetry is guaranteed by the original gauge symmetry even after its breaking to match phenomenology) and some implications of these models. In particular, it is noted that in minimal theories that extend MSSM to include local $B-L$ symmetry, R-parity breaking is mandatory. To the extent that neutrino masses seem to suggest that indeed physics beyond the standard model must include a broken local $B-L$ symmetry, the fact that its supersymmetrization leads to RPV theories, makes a compelling case for theoretically exploring and experimentally testing the idea of R-parity violation. 

To start with, let us note that there are two classes of models of R-parity breaking(RPV): (i) spontaneous breaking by giving vevs to neutral superpartners, which in the case of MSSM is the superpartner of the neutrino (or sneutrino)~\cite{aulakh} or (ii) by adding R-parity violating terms in the superpotential~\cite{hs}. Below I present some models and results pertaining to both these lines of thinking. As we will see below, in case (i), generically, a single vacuum expectation value characterizes all R-parity violating interactions and therefore the theory is quite predictive in RPV sector.  

This article is organized as follows: in sec. 2, I discuss explicit R-parity breaking in MSSM; in sec. 3, I review works on models with global $B-L$ symmetry and in sec. 4, those with local B-L;  in sec. 5, I discuss ways to obtain RPV superpotentials and their implications. In sec. 6, some cosmological implications of RPV are discussed; in sec 7, status of R-parity violation in grand unified theories are noted. Sec. 8, gives some concluding remarks.

\section{Explicit RPV models} In this section we focus on explicit RPV terms in MSSM~\cite{hs}, which can be written as:
\begin{eqnarray}
W_{RPV}~=~\lambda_{ijk}L_iL_je^c_K+\lambda^\prime_{ijk}Q_iL_j d^c_k+\lambda^{\prime\prime}_{ijk}u^c_id^c_jd^c_k +m_iL_iH_u
\end{eqnarray}
This is a class of most widely studied model (see \cite{rev} for a review). These models have 45 RPV coupling parameters. The last term (i.e. the bilinear RPV term) could always be rotated by redefining the the down type Higgs $H_d$ which then simply redefined the $\lambda_{ijk}$ and $\lambda^\prime_{ijk}$ couplings.

First point to note is that if all three i.e. $(\lambda, \lambda^\prime, \lambda^{\prime\prime}$) are present, exchange of squarks will lead to proton decay with amplitudes proportional to $\frac{\lambda^{\prime\prime}}{M^2_{\tilde{d}}}$. Present bounds on proton lifetime imposes severe constraints on them e.g. roughly $\lambda^\prime_{ilk}\lambda^{\prime\prime}_{jlm} \leq 10^{-24}$ for couplings with first generation indices and slightly less severe for higher ones arising from CKM mixings. To avoid excessively small couplings therefore, it is useful to search for extensions of MSSM where only either $\lambda^\prime$ or $\lambda^{\prime\prime}$ is present. In the last couple of years, examples of this type have emerged and we give them below \footnote{There are also much less severe constraints on the individual couplings $\lambda^\prime_{111}$ ~\cite{klap,others} and $\lambda^\prime_{131}$ couplings~\cite{babu} from lepton number violating processes such as $\beta\beta_{0\nu}$ decay.}.

A question has always remained in the explicit RPV models as to how naturally to guarantee proton stability and yet have RPV interactions so that alternative interpretation of LHC limits on SUSY can be given. There are two possibilities: 

\noindent(i) Include baryon parity as an exact symmetry under which all quark superfields change sign and others do not. As a result,  one gets $\lambda^{\prime\prime}=0$ but the other RPV terms remain. This model does not lead to proton decay or any other form of baryon number violation. This is similar to spontaneous RPV models.

\noindent(ii) Look for symmetries that allow the $\lambda^{\prime\prime}$ terms but not $\lambda^\prime$ terms. Again in this case, proton decay is suppressed in leading orders. An example of this kind of model which is motivated by attempts to understand flavor puzzle of SM, is given below.

One class of models where the proton decay problem is automatically avoided are the ones where R-parity is spontaneously broken. We start with them first.

\section{Spontaneous R-parity breaking with global lepton number} The basic idea of this class of models is to start with a theory which conserves R-parity in the Lagrangian and let the vacuum state break it . That would mean giving vacuum expectation value to a superpartner field which is R-parity odd. The relevant fields must of course be color and electrically neutral. In the case of MSSM that leaves only fields that carry a non-zero lepton number as the three sneutrinos. In the extended models there could be more choices. We give two examples below:

\subsection{MSSM example}
 Since we do not want to break electric charge, pretty much the three fields of MSSM which are candidates for acquiring vevs, are the three superpartners of the neutrinos i.e. $\tilde{\nu}_{e,\mu,\tau}$. The first spontaneous RPV model that was written down in ref. ~\cite{aulakh}, used one of these fields and  spontaneous R-parity violation was implemented by giving the electron type sneutrino a non-zero vev. It is clear from the above definition of R-parity that since this field has $S=0$ and $B=0$ and $L=1$, it is odd under R-parity and its vev will therefore break R-parity. Since $\tilde{\nu}_e$ carries lepton number but no baryon number, this theory leads to only lepton number violating but baryon number conserving couplings in the resuting Lagrangian with many phenomenological consequences. From Eq. (1), it is clear that $\tilde{\nu}_{e}$ vev will mix the lepton with the Higgsino. When the mass terms for the lepton fields are diagonalized, it leads to lepton number violation since the Higgsino has no lepton number in the starting Lagrangian. The strength of lepton number violation is dictated by the magnitude of the $\tilde{\nu}_{e}$ vev. The detailed phenomenology of the model was discussed in ~\cite{aulakh}. In 1982, the information available information on the low energy weak interactions was limited and the model was found to be consistent with all data. This was followed up in several papers~\cite{valle} where vev was assigned to the $\tilde{\nu}_{\tau}$ field so that some of the constraints of the previous model ~\cite{aulakh} were weakened . However both these types of models were ruled out by subsequent observations. The main reason is as follows:%It was pointed out in ~\cite{aulakh} that since $<\tilde{\nu}_e>$ has lepton number $L=1$, it breaks R-parity and leads to many lepton number violating signatures. 
since vev of a sneutrino breaks global $B-L$ symmetry spontaneously,  the model predicts a zero mass Goldstone boson, the majoron~\cite{cmp}. The majoron is  the field, $\chi\equiv$ Im$\tilde{\nu}_e$ and is  part of an $SU(2)_L$ doublet field. As a result, it couples to the $Z$ boson together with the Re$\tilde{\nu}_e\equiv \sigma$. So unless the Re$\tilde{\nu}_e$ is heavier than 90 GeV, $Z$ will decay via the $Z\to \chi\sigma$ mode resulting from spontaneous R-parity breaking. This mode will contribute like half a neutrino to the invisible width of the $Z$. $Z$-width was measured very precisely at LEP and SLC and does not have any room for such a large contribution to $Z$ invisible decay. Thus this class of spontaneous RPV models~\cite{aulakh,valle}, where $\tilde{\nu}_\ell$ field has vev and the real part of the $\tilde{\nu}_{\ell}$ field is also light (as it happens in simple versions of the model) , are now ruled out by experiments.

\subsection{Spontaneous RPV with singlet majoron} To avoid this conflict with $Z$ width measurement, a new class of spontaneous RPV models were proposed in ~\cite{masiero} where instead of the left handed sneutrino fields having  a vev, the MSSM was extended to include a right handed neutrino, $N$ and the superpartner of the right handed neutrino field, $\tilde{N}$ was given a vev.  The starting superpotential for this model is:
\begin{eqnarray}
W_N~=~W_{MSSM}+h_\nu LH_uN^c+SN^c\Phi+\Phi^3+\Phi H_uH_d
\end{eqnarray}
The $\Phi$ field has zero lepton number and $N^c$ and $S$ fields have $L=\mp1$ respectively.
In this case, the massless majoron field, corresponding to spontaneous breaking of global lepton number symmetry, is given by linear combination of Im$\tilde{N}$ and Im$S$ fields which are both $SU(2)_L$ singlet fields and the majoron therefore does not couple to the Z-boson. The Z-width constraint therefore does not apply. We note however that generation $\tilde{N}$ vev in this model is not mandatory but is chosen to illustrate the idea. 

In both these classes of models, R-parity of the basic Lagrangian is an assumption as is the vev of the sneutrino field. Below we discuss models where  R-parity becomes an automatic symmetry due to local $B-L$ symmetry when the SM gauge group is extended and in a subclass of these models, $\tilde{N}$ vev is mandated by other physical considerations. We discuss examples of these models below.

\section{Minimal local $B-L$ models and R-parity}
The introduction of local $B-L$ symmetry imposes interesting constraints on R-parity. The first interesting point is that unlike R-parity conserving MSSM  where RPV terms of the superpotential are set to zero by hand, the local $B-L$ symmetry forbids the presence of all RPV terms automatically~\cite{RNM}. This guarantees the stability of neutralino dark matter and provides a more compelling picture of dark matter compared to MSSM. However, whether R-parity ultimately remains a good symmetry or not depends on the detailed structure of the model.  

\subsection{Minimal models with automatic R-parity breaking}
 The gauge group for these models~\cite{RNM1,RNM2,RNM3} is $SU(2)_L\times U(1)_{I_{3R}}\times U(1)_{B-L}$ with all particle assignments given according to their $I_{3,R}, B, L$ quantum numbers. The only extra superfields of the model beyond those of MSSM are $N^c$ (three for three generations to cancel gauge anomalies). The Higgs superfields of the model are the two MSSM doublets $H_u, H_d$.
The superpotential is given by:
\begin{eqnarray}
W_{BL}~=~W_{MSSM}+h_\nu LH_uN^c
\end{eqnarray}
 Therefore to break the local $B-L$ symmetry so that the model becomes phenomenologically viable,  the right handed sneutrino $\tilde{N^c}$ must acquire a vev. Since this field is $RP$ odd, it breaks R-parity; thus R-parity is spontaneously broken. In some sense, one could call these models the ``genuine spontaneous RPV models" since, R-parity violation is mandatory to reduce the gauge symmetry down to the SM gauge symmetry. The simplest complete model of this type is the model of ref.\cite{RNM3}, which uses a minimal set of fields and shows that R-parity is dynamically broken by radiative corrections~\cite{perez1} by the same mechanism that was used for radiative breaking of electroweak symmetry  in minimal supergravity theories of MSSM~\cite{wise}. The massless majoron field which arose in the case of global $B-L$ breaking is now absorbed as the longitudinal mode of the $B-L$ gauge field. Phenomenology of these models, including its LHC signatures, have been studied extensively in ~\cite{RNM3,perez} and are not discussed here. The spontaneous breaking can also be shown to arise radiatively by extrapolating the $\tilde{N^c}$ mass from the GUT scale, which is an interesting feature. It is also worth noting that a gauged B-L 
can have a mass growth for the vector boson preserving R parity within a Stueckelberg mechanism~\cite{nath}

Below the $<\tilde{N^c}>$ scale, the effective MSSM that emerges has the usual RP conserving superpotential together with an RPV term of the form $h_{\nu, ij}<\tilde{N^c}_j>L_iH_u$. This is a bilinear RPV term~\cite{romao} and by rotation of the $H_d,L$ fields it can be removed and replaced by the  RPV superpotential where the couplings of the explicit RPV terms  are predicted in terms of quark  and lepton Yukawa couplings and the ratio $h_{\nu, ij}<\tilde{N^c}_j>/\mu$ where $\mu$ is the familiar $\mu$-term of MSSM. 
It is also clear that it induces only only terms of type $LLe^c$ and $QLd^c$ and thus no baryon number violation. This is an interesting feature of such models which enhances the appeal of such theories for RPV. Neutrino masses in these models arise from the inverse seesaw ~\cite{RNM1,MV} mechanism and have an interesting ``layered" pattern that includes sterile neutrinos, as noted in ref.~\cite{goran}. 

%and as an optional additiona pair field $\Delta^0$ with $B-L=2$ to implement the seesaw mechanism

\subsection{ From RPV to RPC with local $B-L$ models}
%TeV left-right seesaw model with special texture from symmetries}\label{sec2}
A slight extension of the above minimal models enables the R-parity to be an exact symmetry. To do this,  we need to include one pair of additional Higgs fields with $B-L=2$, called $\Delta^0$ and $\bar{\Delta}^0$. The new superpotential for these models is:
\begin{eqnarray}
W_{seesaw}~=~W_{BL}+f N^cN^c\Delta^0+\lambda X( \Delta^0\bar{\Delta}^0-M^2)
\end{eqnarray}
The ground state of this theory corresponds to $<\Delta^0>=<\bar{\Delta}^0>=M$ as can be seen by minimizing the F- and D-term contributions to the potential. This breaks $B-L$ by two units and therefore keeps R-parity exact. It also gives mass to the right handed neutrinos so that seesaw formula (type I)~\cite{seesaw}  for neutrino masses i.e. $m_\nu~=~-\frac{(h^2_{\nu}v^2_{wk})}{fM}$ emerges. For  $h_\nu\sim 10^{-5}$ and $M\sim 10$ TeV, this gives right order of magnitude for the neutrino masses in neutrino oscillation experiments.

\subsection{Left-right embedding of local $B-L$ and R-parity} Something interesting happens, once  the model with local $B-L$ (subsection B above) is embedded into its left-right symmetric version and the $X$-field is omitted. To explain this, let us start with a brief overview of the left-right models.

 In the minimal L-R model, the fermions are assigned to the gauge group $SU(2)_L\times SU(2)_R\times U(1)_{B-L}$ as follows: denoting
%$Q\equiv\left(\begin{array}{c}u\\d\end{array}\right)$ and $\psi\equiv\left(\begin{array}{c}\nu\\e\end{array}\right)$
$Q\equiv (u,d)^{\sf T}$ and $L\equiv (\nu_\ell, \ell)^{\sf T}$ as the quark and lepton doublets respectively, $Q$ and $L$  are assigned to doublets under the $SU(2)_L$ group, while $Q^c$ and $L^c$  as the doublets under the $SU(2)_R$ group~\cite{LR}. Their $B-L$ quantum numbers can easily be worked out from the definition of the electric charge: $Q=T_{3L}+T_{3R}+(B-L)/2$, where $T_{3L}$ and $T_{3R}$ are the third components of isospin under $SU(2)_L$ and $SU(2)_R$ respectively.
%expected quantum numbers.
The Higgs sector of the model consists of one or several of the following multiplets:
\begin{eqnarray}
\Delta_R\equiv\left(\begin{array}{cc}\Delta^+_R/\sqrt{2} & \Delta^{++}_R\\\Delta^0_R & -\Delta^+_R/\sqrt{2}\end{array}\right),~ \phi\equiv\left(\begin{array}{cc}\phi^0_1 & \phi^+_2\\\phi^-_1 & \phi^0_2\end{array}\right)
\end{eqnarray}
The $\Delta_R$ superfield must be accompnaied by a $\bar{\Delta}_R$ field to cancel anomalies. Al;so note that the bi-doublet field contains two SM doublets. In the language of MSSM, it contains the $H_u,H_d$ of MSSM unified in their couplings by parity. For example, each $\phi$ coupling to quarks leads to both down and up Higgs couplings to be equal. As a result to get CKM mixings, we need at least two bi-doublets ib SUSYLR models of this type.
The gauge symmetry $SU(2)_R\times U(1)_{B-L}$ is broken by the vev $\langle \Delta^0_R\rangle = v_R$ to the group $U(1)_Y$ of the SM.
There is also an LH counterpart ($\Delta_L$) to $\Delta_R$ which we do not consider here.  There are versions of the model where parity and $SU(2)_R$ gauge symmetry scales are decoupled so that the $\Delta_L$ fields become heavy when the discrete parity symmetry is broken. The low energy Lagrangian in this case has invariance under the L-R gauge group but not parity. We will focus on this class of models in this paper, since they seem to be sufficient to illustrate our points regarding RPV. %Left-right symmetry implies that the SM Higgs responsible for giving mass to the fermions and the $W_L$ and $Z$ boson is part of a single or  a set of bi-doublet fields  $\phi_a\equiv\left(\begin{array}{cc}\phi^0_1 & \phi^+_2\\\phi^-_1 & \phi^0_2\end{array}\right)$.
The vev of the $\phi$ field given by $\langle\phi\rangle={\rm diag}(\kappa, \kappa')$ breaks the SM gauge group to $U(1)_{\rm em}$.
The superpotential for this model is given by:
\begin{eqnarray}
{W}_{LR}&=&h^{q,a}_{ij}{Q}_{i}\phi_aQ^c_{j}+
h^{\ell,a}_{ij}{L}_i\phi_aL^c_j 
%&&+ \tilde{h}^{\ell,a}_{ij}\bar{L}_i\tilde{\phi}_aR_j
%+h^a_{q,ij}\bar{Q}_{L,i}\phi_aQ_{R,j} +\tilde{h}^a_{q,ij}\bar{Q}_{L,i}\tilde{\phi}_aQ_{R,j}+%\\ %\nonumber
+f_{ij} L^c_iL^c_j\Delta_R+\mu {\rm Tr} \phi_a\phi_b +\mu_\Delta \Delta_R\bar{\Delta}_R
\label{eq:yuk}
\end{eqnarray}
where $i,j$ stand for generations and $a$ for labeling the Higgs bi-doublets, and $\tilde{\phi}=\tau_2\phi^*\tau_2$ ($\tau_2$ being the second Pauli matrix). After symmetry breaking, the Dirac fermion masses are given by the generic formula $M_f~=~h^f\kappa + \tilde{h}^f\kappa'$ for up-type fermions, and for down-type quarks and charged leptons, it is the same formula with $\kappa$ and $\kappa'$ interchanged.  The Yukawa Lagrangian (\ref{eq:yuk}) leads to the Dirac mass matrix  for neutrinos $M_D = h^{\ell}\kappa + \tilde{h^{\ell}}\kappa'$ and the Majorana mass matrix for the heavy RH  neutrinos $M_N=fv_R$ which go into the equation for calculating the neutrino masses and the heavy-light neutrino mixing.

It was shown in ref.~\cite{kuchi} that  in this minimal susy left-right seesaw model, if we express the scalar potential, $V$ as functions of the vevs of the neutral Higgs fields ( $v_R$ , $\bar{ v}_R$ and $x\equiv <\tilde{N^c}>$ and look for a minimum of $V$ along the direction $x = 0$ (i.e. R-parity conserving ), the minimum occurs at $ v_R = \bar{v}_R = 0$. We then show that once we include R-parity breaking effect by  the vacuum i.e. $<\tilde{N^c}> = x\neq 0$ i.e. along the direction $x\neq 0$, there appear global minima that break
parity i.e. $v_R,v_R\neq 0$ and also that it occurs only below a certain value for $v_R$ and $\bar{v}_R$ i.e. there is an upper limit on the parity breaking scale. Thus, breaking of R-parity is dynamically induced in the correct vacuum. No choice of parameters is required for this result. This is an example of dynamical breaking of R-parity~\cite{zhang}\footnote{For other examples of dynamical RPV, see ~\cite{csaki}.}.

In fact, in this minimal left-right seesaw model, the requirement of gauge symmetry breaking leads to an upper limit on the right handed $W_R$ scale of $M_{W_R}\leq \frac{gM_{SUSY}}{f}$ ~\cite{kuchi2} where $f$ is an average value of the Majorana Yukawa coupling (defined in Eq.~\ref{eq:yuk}).

This result led to further investigation to see under what conditions a left-right seesaw model can lead to an R-parity conserving vacuum~\cite{RPC, RPC1}. This generally involves adding higher dimensional terms to the superpotential or new fields  e.g.  triplet Higgs fields with $B-L=0$~\cite{kuchi,RPC} with non-zero vevs or for some domain of parameters, simply including one loop radiative corrections~\cite{RPC1}.

\section{Gauged Flavor and RPV pattern}
There are many attempts in the literature to understand the flavor problem. One promising direction starts from the realization that in the limit of vanishing Yukawas of the standard model, the family (flavor) symmetry of the model becomes $[SU(3)]^5$. The hope then is that breaking of this symmetry could provide an understanding of the flavor pattern of fermions. The question then is whether this is a global or local symmetry. Regardless of this question, one could discuss the implications of such models for RPV as has been done in refs.~\cite{yuval,roberto}.

The global version will be plagued by the appearance of massless states which will affect many aspects of cosmology. Attempts therefore have been made to pursue these symmetries as local symmetries~\cite{GRV,gms}. This requires the introduction of vector-like SM singlet fermions. Once this model is supersymmetrized, it has interesting consequences for R-parity that we discuss below~\cite{roberto}.
 
We focus on supersymmetric versions of the gauged flavor with left-right symmetric electroweak interactions  Ref. \cite{gms}  {from which we borrow the notation.} %$U(3) 
%Here we consider the supersymmetric version of this model. %Some aspects of this model were noted in Ref.~\cite{GFRPV}.
The largest flavor group for this case is $SU(3)_{Q_L} \times SU(3)_{Q_R}\times SU(3)_{\ell_L}\times
SU(3)_{\ell_R}$.
{For the cancellation of anomalies we introduce vector-like superfields $U, U^c,D, D^c$ and $E, E^c,N, N^c$, analogously to the case of the MSSM.}
{The overall anomaly free gauge group is therefore}  $$G_{LR} \equiv
SU(3)_{c} \times SU(2)_L \times SU(2)_R \times U(1)_{B-L} \times
SU(3)_{Q_L} \times SU(3)_{Q_R}\times SU(3)_{\ell_L}\times
SU(3)_{\ell_R},$$ where $SU(3)_{Q_L} \times SU(3)_{Q_R}$
represents the flavor gauge symmetries respectively in
the left- and right-handed quark sector, and $SU(3)_{\ell_L}
\times SU(3)_{\ell_R}$ the corresponding ones for the lepton sector.
{The electroweak part of the SM gauge group is embedded into the group $SU(2)_{L}\times SU(2)_{R}\times U(1)_{B-L}$ that is broken to the SM group $SU(2)_{L}\times U(1)_{Y}$ by the VEV $v_{R}$ of a set of fields $\chi^{c},\bar{\chi}^{c}$.}
{The flavor symmetry is broken by a set of flavons $Y_{u},\bar{Y}_{u},Y_{d},\bar{Y}_{d}$ and $Y_{\nu},\bar{Y}_{\nu},Y_{\ell},\bar{Y}_{\ell}$ as in the example of the MSSM}.
 The fields of the model and their transformation properties under
 the group $G_{LR}$ are reported in Table \ref{tab:modelone}. 
 \begin{table}
\small
\begin{center} \begin{tabular}{|| l | c  c c c  | c c c c || }\hline\hline
&  $SU(3)_c$ & $SU(2)_L$ & $SU(2)_R$ & $U(1)_{B-L}$ & $SU(3)_{Q_L}$ &
  $SU(3)_{Q_R}$ &  $SU(3)_{\ell_L}$ & $SU(3)_{\ell_R}$ \\ [3pt]
\hline \hline
  $Q$ & 3 & 2 & & $\frac{1}{3}$ & 3 &    & & \\ [3pt]
  $Q^c$ & $3^*$ & & 2 & $-\frac{1}{3}$ &   & $3^*$  & & \\ [3pt]
   $U$ & 3 & & & $\frac{4}{3}$   &   & 3  & & \\ [3pt]
    $U^c$ & $3^{*}$ & & & $-\frac{4}{3}$   & $3^{*}$ &    & & \\ [3pt]
   $D$ & 3 & & & $-\frac{2}{3}$ &    & 3  & & \\ [3pt]
    $D^c$ & $3^*$ & & & $\frac{2}{3}$  & $3^*$ &   & & \\ [3pt]
\hline
  $L$ & & 2 & & $-1$ & &  & 3 &   \\ [3pt]
  $L^c$ && & 2 & $1$ & & &   & $3^*$ \\ [3pt]
     $E$ & &&   & $-2$ & & &   & $3$ \\ [3pt]
     $E^c$ && &   & $2$ &  & & $3^{*}$ &   \\ [3pt]
     $N$ && &   & 0  & & &    & 3 \\ [3pt]
     $N^c$ && &   & 0  & &  & $3^*$ &   \\ [3pt]
\hline
  $\chi,\bar{\chi}$ & & 2 & & $\pm 1$ &  &  & &\\ [3pt]
  $\chi^c,\bar{\chi}^c$ & & & 2 & $\pm 1$ &  &  & & \\ [3pt]
  $Y_u$ & &  &  &  & $3$ & $3^{*}$ &&\\ [3pt]
  $\bar{Y}_u$& &  &  &  & $ {3^{*}}$ &$3$ && \\ [3pt]
  $Y_d$ &  &&  &  & $3$ &  $3^{*}$  & &\\ [3pt]
   $\bar{Y}_d$ & &  &  &  & ${3^{*}}$ & $3$  &  &\\ [3pt]
  $Y_\ell$ & & & & & & &$3$ &  $3^{*}$\\[3pt]
  $\bar{Y}_\ell$ & & & & & & &${3^{*}}$ & $3$\\[3pt]
$Y_\nu$ & & & & & & &$3$ &  $3^{*}$\\[3pt]
$\bar{Y}_\nu$ & & & & & & &${3^{*}}$ & $3$\\[3pt]
\hline
\end{tabular} \end{center}
\caption{Model I matter content and transformation properties. The MSSM fields are the lines with green background, the new matter fields are given in the yellow lines, and the electroweak and flavor symmetry breaking sector is given in the lines with white background.}
\label{tab:modelone}
\end{table}
\normalsize

The interaction of the MSSM  and the exotic fields is given by the superpotential 
$$W_{I} = \lambda_{u}( Q\chi U^{c} + Q^{c}\bar{\chi}^{c} U) + \lambda_{d}( Q\bar{\chi} D^{c} + Q^{c}\chi^{c} D) + \lambda_{u}^{\prime}Y_{u}UU^{c} + \lambda_{d}^{\prime}Y_{d}DD^{c}\,, $$
where the equality of the couplings are dictated by unbroken L-R parity at the scale at which we write this superpotential. {This assumption about the L-R parity can be removed without affecting our conclusions, though getting more involved formulae. Thus we will discuss explicitly only the L-R parity symmetric case.}
{From these interactions the mass terms are generated once the flavon fields $Y$ and the Higgs fields $\chi$ take a VEV. To fix our notation we take $<{\chi^{c}}>=<{\bar{\chi}^{c}}>=(0,v_{R})$ and $<{\chi}>=<{\bar{\chi}}>=(0,v_{L})$.
The flavons can be written in a basis where the $Y_{d}$ are diagonal by mean of a suitable flavor rotation, after which the $Y_{u}$ is fixed. Therefore we assume that the VEV of the flavons are such that  $<{Y_{d}}>=<{\bar{Y}_{d}}>=<{\hat{Y}_{d}}>$ and $<{Y_{u}}>=<{\bar{Y}_{u}}>=V_{CKM}<{\hat{Y}_{u}}>V_{CKM}^{\dagger}$ where hat denotes diagonal matrixes.} 

As shown in  Ref.~\cite{gms} in this model the SM fermion masses are given by  a seesaw formula: 
\begin{eqnarray}
M_d\simeq {\lambda^2_d \bar{v}_L v_R \over \lambda_{d}^{\prime}} \langle\hat{Y}_d\rangle^{-1},\quad M_u\simeq  {\lambda^2_u v_L v_R \over \lambda^{\prime}_{u}} \cdot V^{\dagger}_{CKM} \langle\hat{Y}_u\rangle^{-1}V_{CKM}\,.
\end{eqnarray}
%\footnote{\red{Here $v_L$ and $\bar{v}_L$ are the analogs of the vevs of $H_u$ and $H_d$ in MSSM and therefore their ratio gives the parameter tan$\beta$. The factor tan$\beta$ can be absorbed into the redefinition of $\lambda^\prime_d$ and we therefore do not display it below.}}.
As in the model discussed above, this implies a mass hierarchy among the vector-like states inverted with respect to that of the SM states. Therefore  the top ``partner'', and possibly its SUSY partner, is predicted to be the lightest colored exotic fermion with masses in the TeV range depending on the scale of $SU(2)_R$ breaking.
{In general the light fermions are an admixture of the usual MSSM interaction eigenstates and the exotic fields with mixing angle that, in the approximation $v_{L} \ll v_{R} \ll <{Y}>$, reads 
$$\theta_{L,R}^{\,(u,d)} \simeq \frac{\lambda_{u,d} \,v_{L,R}}{\lambda_{u,d}^{\prime}<{Y_{u,d}}> }\,. $$}

{The RPV couplings in this model are given by renormalizable interactions among the exotic states:}
\begin{eqnarray}\label{RP}
W_{RPV}~=~\lambda_q\epsilon_{ijk} \left[ U^c_iD^c_jD^c_k+U_iD_jD_k\right]+\lambda_\ell \epsilon_{ijk} \left[ N^iN^jN^k+N^{c,i}N^{c,j}N^{c,k}\right]\,, \label{RPVmodelI}
\end{eqnarray}
where the explicit indexes are flavor indexes of the relevant flavor gauge group. The RPV couplings for the light states originate from the mixing of the interaction eigenstates.
%In order to get the RPV couplings for light SM fermions (and in general for the MSSM light matter superfields), we  start with a basis where \blue{with a $SU(3)_{Q_{L}}$ and a $SU(3)_{Q_{R}}$ gauge transformation we make} the down quarks masses  diagonal (since we choose $\langle Y_{d}\rangle$ diagonal). On the other hand there is the CKM mixing factor in the up quark sector.  
The three RPV couplings involving the vector-like quarks are $U^c_1D^c_2D^c_3$, $U^c_2D^c_1D^c_3$, $U^c_3D^c_1D^c_2$ and taking account of the mixings, we get the effective $\Delta B=1$ R-parity violating couplings in the Table~\ref{tab:couplingsmodel1}. {In the Table, we have assumed the scalar quark mass matrix to be diagonalized by the same rotation as the fermionic one~\footnote{{Deviations from this assumption are expected when one considers SUSY breaking effects. However, to include this realistic element of the description of the RPV interaction one needs to specify the mechanism of SUSY breaking mediation, which we leave for future work.}}, also we have omitted a factor $\lambda_q/(\lambda_d^2\lambda_{u})$ that is common to all the coupling and that we can take to be of order one.}  The reason for the appearance of only one CKM factor is that we have chosen a basis where the down sector is flavor diagonal to start with and all CKM factors come from the up-sector. In the up sector, we get $U^c=V_{CKM} U^c$ and the mixing between heavy and light quarks is given by: $V^{\dagger}_{CKM}\frac{\lambda_u}{\hat{Y}_u}$.

\begin{table}
%\small
\begin{center} \begin{tabular}{||c|c|| }\hline
$\Delta B=1$ operator & strength \\[3pt]\hline
$u^cs^cb^c$ &$ {V_{ud} \, m_u\,m_s\,m_b}/{m^3_t}$\\[3pt]
$c^cs^cb^c$ &$ {V_{us}\, m_c\,m_s\,m_b}/{m^3_t}$\\[3pt]
$t^cs^cb^c$ &$ {V_{ub}\, m_t\,m_s\,m_b}/{m^3_t}$\\[3pt]
$u^cd^cb^c$ &$ {V_{cd}\, m_u\,m_d\,m_b}/{m^3_t}$\\[3pt]
$c^cd^cb^c$ &$ {V_{cs}\, m_c\,m_d\,m_b}/{m^3_t}$\\[3pt]
$t^cd^cb^c$ &$ {V_{cb}\, m_t\,m_d\,m_b}/{m^3_t}$\\[3pt]
$ u^cd^cs^c$ &$ {V_{td}\, m_u\,m_d\,m_s}/{m^3_t}$\\[3pt]
$ c^cd^cs^c$ &$ {V_{ts} \,m_c\,m_d\,m_s}/{m^3_t}$\\[3pt]
$ t^cd^cs^c$ &$ {V_{tb}\, m_t\,m_d\,m_s}/{m^3_t}$\\[3pt]
%$u^cs^cb^c$ &$ \frac{V_{ud} m_um_sm_b}{m^3_t}$\\[3pt]
%$c^cs^cb^c$ &$ \frac{V_{us} m_cm_sm_b}{m^3_t}$\\[3pt]
%$t^cs^cb^c$ &$ \frac{V_{ub} m_tm_sm_b}{m^3_t}$\\[3pt]
%$u^cd^cb^c$ &$ \frac{V_{cd} m_um_dm_b}{m^3_t}$\\[3pt]
%$c^cd^cb^c$ &$ \frac{V_{cs} m_cm_dm_b}{m^3_t}$\\[3pt]
%$t^cd^cb^c$ &$ \frac{V_{cb} m_tm_dm_b}{m^3_t}$\\[3pt]
%$ u^cd^cs^c$ &$ \frac{V_{td} m_um_dm_s}{m^3_t}$\\[3pt]
%$ c^cd^cs^c$ &$ \frac{V_{ts} m_cm_dm_s}{m^3_t}$\\[3pt]
%$ t^cd^cs^c$ &$ \frac{V_{tb} m_tm_dm_s}{m^3_t}$\\[3pt]
\hline
\end{tabular} \end{center}
\caption{Predictions for the $\Delta B=1$ RPV couplings in the gauged flavor model I. The strength of the coupling is given up to a factor $\lambda_q/(\lambda_d^2\lambda_{u})$ that is a parameter of the model and can be order one.}
\label{tab:couplingsmodel1}
\end{table}

We thus see that beyond the standard model with with gauged flavor limits the pattern of RPV in such a way that it avoids the proton decay problem and also leads to observable signals in $\Delta B=2$ transitions ~\cite{roberto} via the Feynman diagram in Fig.1.  In the leading order this diagram leads to $\Delta B=2$ processes ~\cite{BLV} di-proton decay $pp\to K^+K^+$ on whose life time there is a lower limit of $1.7\times 10^{32}$ yrs~\cite{miura}. In conjunction with a $\Delta S=2$ or $\Delta b=2$ ($b$ standing for bottom flavor) interactions, such diagrams can also lead to  neutron-anti-neutron oscillation~\cite{goity}.
 \begin{figure}[h!]
\centering
\includegraphics[scale=0.8]{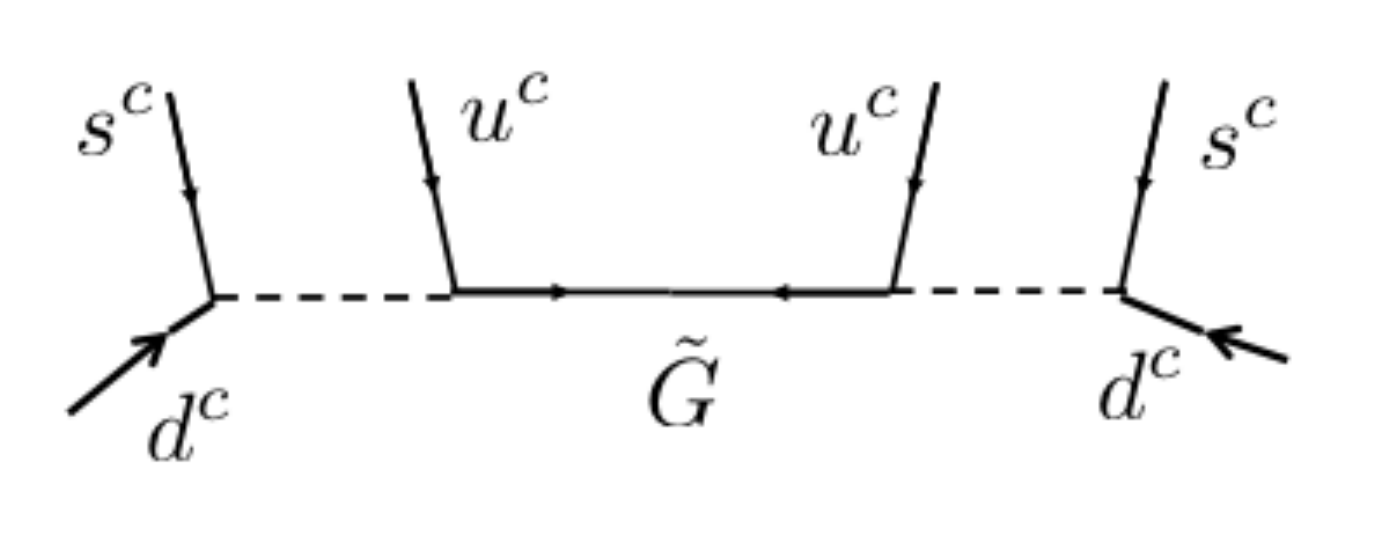}
\caption{tree diagram for $\Delta B=2$ process induced by $\Delta B=1$ RPV terms}
\label{radseesaw}
\end{figure}

\section{Cosmological implications} In this section, we emphasize two cosmological implications of RPV interactions: oner for the origin of matter and second for the nature of dark matter of the universe.

\subsection{Implications for baryon asymmetry}
The strengths of RPV interactions are constrained once we assume that matter-anti-matter asymmetry arises above the supersymmetry breaking scale i.e. above the TeV scale. The point is that since the RPV interactions break baryon and lepton number, unless their couplings are very small, those interactions will be in equilibrium and erase any baryon or lepton  number asymmetry introduced in the earlier epochs. The typical requirement is that the rate for $\Delta B\neq 0$ or $\Delta L\neq 0$ interactions must be slower than the expansion rate of the universe i.e.
\begin{eqnarray}
\Gamma_{RPV}\leq {g^*}^{1/2}\frac{T^2}{M_{P\ell}}
\end{eqnarray}
This gives typically that $\lambda^2 \leq \frac{100 {\rm TeV}}{M_{P\ell}}\leq 10^{-14}$~\cite{david} and implies that the RPV couplings are typically less than $10^{-7}$ in their strength, if they are not to erase any pre-existing baryon or lepton asymmetry of the universe. It is interesting that the RPV predictions of the gauged flavor model in Table II are consistent with these bounds. Of course, it could be that baryon or lepton asymmetry is generated below 10 TeV scale and if we assume that all superpartner masses are above 10 TeV, then such constraints do not apply. Furthermore, it has also been speculated that RPV itself may be a source of the cosmological matter-anti-matter asymmetry which arises in the TeV scale~\cite{raby} via the decay of gravitino using RPV modes.

The other possibility is that spontaneous RPV scale is in the TeV range so that above that scale, R-parity becomes an exact symmetry and provides a way to avoid these constraints.

\subsection{Dark matter and RPV} As eluded to above, in the presence of R-parity conservation, the lightest super-partner is stable and can be assumed to be the dark matter (DM) of the universe. The conventional choice for DM is the neutralino, which is a Majorana fermion and is a linear combination of the Wino, Bino and the two neutral Higgsinos. However, once R-parity breaking is included in the Lagrangian, neutralino becomes unstable and cannot anymore play the role of dark matter. However, supergravity theories can then provide another particle which can be the dark matter i.e. heavy gravitino, the superpartner of the graviton. In the presence of RPV interactions, the gravitino is unstable and can decay. However, since its presence is connected to gravity, its decay strength is suppressed by the gravitational coupling i.e. it coupling to matter is roughly given by $\frac{\kappa_i}{M_{Pl} }$ where $\kappa_i$ is related to the strength of R-parity breaking~\cite{decay}.  This gives its lifetime as $\tau_{\tilde{g}}\sim \frac{192\pi^3 M^2_{Pl}}{\kappa^2_im^3_{\tilde{g}}}$. For $\kappa_i\sim 10^{-7}$, $\tau_{\tilde{g}}$ can be much longer than the age of the universe. Often in actual UV complete models this can be even longer. Thus gravitino can play the role of a decaying dark matter~\cite{decay}. While true nature of dark matter to date remains a mystery, there are lepton signals in various experiments which have been interpreted in terms of an unstable gravitino dark matter~\cite{decay}. Thus in an ironic sense, it is not R-parity conservation but rather R-parity breaking which may be relevant to dark matter issues if these signals and others accumulate over time.

\section{R-parity and grand unification} In this section, we discuss the status of R-parity in supersymmetric grand unified theories such as SU(5) and SO(10). 

\subsection{SU(5)}We briefly recall that in the SUSY $SU(5)$ model, the matter multiplets belong to {\bf 10} and $ \bar{5}$ representations. It is then clear that one can construct renormalizable interactions involving them e.g. $W^\prime\equiv {\bf 10}\cdot { \bar{5}}\cdot {\bar{5}}$. Since  the matter content of $ (Q, u^c,e^c)\in {\bf 10}$ and $ (d^c, L)\in {\bar{5}}$, it follows that $W^\prime$ leads to all three RPV terms of MSSM i.e. $u^cd^cd^c$, $QLd^c$ and $LLe^c$. For some implications of these terms, see Ref.~\cite{bajc}.

\subsection{SO(10)} The situation is somewhat different in the SO(10) model, which contains local $B-L$ as a sub-symmetry. Therefore as in the left-right case discussed above, prior to symmetry breaking, R-parity is an exact symmetry.  However, since the original $SO(10)$ symmetry must be broken down to the standard model, whether R-parity is broken in the low energy theory or not depends on how the original gauge symmetry is broken. For instance, depending on whether $SO(10)$ is broken down to SU(5) by (i) {\bf 16}-dimensional Higgs field or (ii) by {\bf 126} field determines whether the low energy theory (or SU(5) in this case) is RPV type (in case (i)) or RPC type (in case (ii))~\cite{dagyu}. Both these classes of models have been discussed in the literature and their various phenomenological implications studied. These two cases exactly parallel the susy left-right model where $B-L$ is broken by a doublet $(1,2,-1)$ or $(1,3,-2)$ Higgs field. 

To illustrate how RPV terms arise in models with $\chi \equiv {\bf 16}_H$ fields, we repeat the point noted above  that in SO(10) models, matter fields belong to also ${\bf 16}_m\equiv \psi$ representations. In the Higgs sector, in addition to $\chi$ fields, there are fields belonging to {\bf 10}-dim. reps (denoted by $H$) and which are needed to break the $SU(2)_L\times U(1)_Y$ symmetry and fields belonging to {\bf 54}-dim. ($S$) and {\bf 45}-dim ($A$). In this case, it is clear if no other symmetries are imposed on the theory, it will allow couplings of the form $\psi \chi H$ as well as other higher dimensional terms e.g. $\psi \psi\psi\chi$ that will break R-parity after symmetry breaking.

%\section{Summary} A brief overview of some theoretical aspects of both spontaneous and explicit R-parity violation is given. 

\section{Summary}
R-parity violation is an integral part of supersymmetry and its existence is mandatory in the simplest supersymmetric theories that contain local $B-L$. Regardless of whether it is explicit or spontaneous, the phenomenon of R-parity violation has many facets to it and touches on such diverse areas of particle physics as low energy rare processes~\cite{rabi}, cosmological baryon asymmetry, LHC signals~~\cite{prashant}, as well as  nature of dark matter. Its signals could throw light even on questions relating to the origin of fermion flavor. It is hoped that this brief overview provides sufficient sense of the excitement and subtleties associated with this line of thinking and will stimulate further progress in this area.

\section*{Acknowledgement} I would like to thank Pran Nath and Pavel Perez for carefully reading the manuscript and providing many helpful comments. This work  is supported by the NSF grant No. PHY-1315155.

%%%%%%%%%%%%%%%%%%%%

\end{document}